\documentstyle[epsfig,pre,multicol,aps]{revtex}

\begin{document}
\begin{title}
{\bf Effects of spin-elastic interactions in frustrated
Heisenberg antiferromagnets}
\end{title}

\author{ Yu. Gaididei\\ 
Bogolyubov Institute for
Theoretical Physics, 252143 Kiev, Ukraine\\
H. B\" uttner \\Physikalishes Institut, Universit\" at Bayreuth,
D95440 Bayreuth, Germany}

\maketitle

\begin{abstract}
The Heisenberg antiferromagnet on a compressible triangular lattice in the
spin-wave approximation is considered.
 It is shown that the interaction between quantum fluctuations and elastic
 degrees of freedom stabilize  the  low symmetric L-phase with a collinear
 N{\'e}el magnetic ordering. Multi-stability in the dependence of the
 on-site magnetization is on an uniaxial pressure is found. 
\end{abstract}
\begin{multicols}{2}
There is a long standing interest in properties of frustrated quantum
antiferromagnets because  they display a large variety of behavior: e.g.
conventional spin ordering, spin liquid ground states, or chirality
transitions \cite{rev}. The antiferromagnetic Heisenberg model on a
triangular lattice is the simplest two-dimensional frustrated system.
Quasi-two-dimensional triangular lattice antiferromagnets are not rare
objects. It was recently shown \cite{bram,serrano} that the materials
$AM(SO_4)_2$ belonging to the Yavapaiite family may
 be considered as realizations of a 
quasi-two-dimensional triangular lattice quantum ({\em e.g.} M=Ti with
S=1/2)
 and quasiclassical
 ({\em e.g.} M=Fe with S=5/2) antiferromagnets. The  rhombohedral
 $\beta$-phase
and monoclinic $\alpha$-phase of solid oxygen consist of weakly
interacting  planes (triangular in the case of $\beta-O_2$ and rectangular
in the case of $\alpha-O_2$ \cite{krup}) whose magnetic properties are
described by the easy-plane Heisenberg antiferromagnetic  model
 with spin $S=1$ \cite{gaid75}. 

Unconventional
properties of the $\kappa-(BEDT-TTF)_2\,X$ family can be described by a
Hubbard model on a triangular lattice with an anisotropic interactions
\cite{mcken}. An antiferromagnetic Heisenberg model was also used in
\cite{rodr} to describe the properties of a triangular monolayer of Pb and
Sn adatoms on the (111) surface of Ge form.

 From the theoretical point of view the antiferromagnetic Heisenberg model
 on a
triangular lattice  has attracted  particular attention since 
Anderson's  proposal of a resonance-valence-bond state for this model
\cite{andfaz}. From that time much work has been done to understand the
nature of its ground state. There is now an almost consensus based on
variety of methods \cite{huse,joli,deu,cap} that the ground state has a
conventional three-sublattice order but the situation is close to marginal
and quantum fluctuations are important \cite{leung}. Quite recently the
effect of quantum spin fluctuations on the ground-state properties of
spatially anisotropic Heisenberg antiferromagnet in the linear spin-wave
approximation was explored \cite{uma,trumpo,merino}. The staggered
magnetization and the magnon dispersion were calculated as functions of
the ratio of the antiferromagnetic exchange between the second and first
neighbors,$~J_2/J_1$. As a physical tool which allows to tune the ratio
$J_2/J_1$ Merino et al \cite{merino} invoke an uniaxial stress within a
layer whereby the relative distances between atoms are changed.

In this paper we investigate the ground-state properties of the
compressible frustrated antiferromagnets, i.e. those systems where the
coupling between magnetic and elastic degrees of freedom is taken into
account,under the action
 of uniaxial pressure. We treat the 
system self-consistently considering the spin subsystem in the linear
spin-wave approximation and  using the continuum approach for the elastic
degrees of freedom. We show that as a result of this method the behavior
of the on-site
 magnetization as a function of the pressure differs significantly from
 the behavior which was obtained in the framework of the spatially
 anisotropic Heisenberg model \cite{trumpo,merino}. In particular, the
 on-site-magnetization does not vanish in the whole interval of stability
 of the collinear phase.

We consider the two-dimensional triangular lattice. The 
Hamiltonian of the spin subsystem, interacting with the displacements of
the magnetic atoms, is \begin{eqnarray} \label{ham} H=\frac{1}{2}\,
\sum_{\vec{n},\vec{a}}J_{\vec{a}}
\vec{S}_{\vec{n}}\vec{S}_{\vec{n}+\vec{a}} \end{eqnarray}

Here ${\vec S}_{\vec{n}}$ is the spin operator of the atom located
in the $\vec{n}$-th site of the triangular lattice: $\vec{n}=n_1
\vec{c}_1+n_2\vec{c}_2\,(n_1,n_2=0,\pm 1,\pm 2,..)$ where
$\vec{c}_1=(1,0),\vec{c}_2=(\frac{1}{2},\frac{\sqrt{3}}{2})$ are the
basic vectors of the triangular lattice,  
$J_{\vec{a}}=J
\left(1-\eta\,\vec{a}(\vec{u}_{\vec{n}+\vec{a}}-
\vec{u}_{\vec{n}})\right)=
J\,\left(1-\eta \sum_{i,j}a_i\,u_{i j} a_j\right)$
is the exchange integral ($ J > 0$ is the exchange constant),
$\vec{u}_{\vec{n}}$ are the atom displacements from their equilibrium
(without magnetic interaction),  and $\vec{a}$
 is  the vector which connects a site with nearest 
neighbors, $\eta=-d \ln J/d |\vec{a}|$.
We  consider a small spatially smooth deformation. 
Therefore it was
convenient  to introduce the components of the strain tensor
$u_{i j}$: $~(\vec{u}_{\vec{n}+\vec{a}}-\vec{u}_{\vec{n}})_i=
\frac{1}{2}\sum_{j}\left(\frac{\partial u_{i}}{\partial
n_j}+\frac{\partial u_{j}}{\partial n_i}\right)\,a_j\equiv \sum_{j}
u_{ij}\,a_j,~$ ($i,j=x,y).$

The elastic energy of the two-dimensional triangular lattice in the
continuum approximation coincides with the elastic energy of the isotropic
plane:$~~\Phi=\frac{1}{2}\,K\,(u_{xx}+u_{yy})^2+\frac{1}{2}\mu\,
\left((u_{xx}-u_{yy})^2+4\,u_{xy}^2\right) -p_x u_{xx}-p_y u_{yy}~~$ where
$K$ and $\mu$ are the elastic modulus of the system and $p_i\,(i=x,y)$ are
the component of the uniaxial pressure.

Let us use the transformation (see {\em e.g.} \cite{tyab}) to the local
frame of reference $S^x_{\vec{n}}=\tilde{S}^x_{\vec{n}} \cos
\vec{q}\vec{n}+ \tilde{S}^z_{\vec{n}} \sin \vec{q}\vec{n},~
S^z_{\vec{n}}=\tilde{S}^z_{\vec{n}} \cos \vec{q}\vec{n}-
\tilde{S}^x_{\vec{n}} \sin \vec{q}\vec{n},
~S^y_{\vec{n}}=\tilde{S}^y_{\vec{n}}$
 in which quantization axis for the spins 
at each site coincide with its classical direction 
$(\cos \vec{q}\vec{n},0,\sin \vec{q}\vec{n})$ which is determined
by the vector $\vec{q}$.
By using the Holstein-Primakoff spin-representation
 $\tilde{S}^x_{\vec{n}}=\sqrt{S/2}
 (b^{\dagger}_{\vec{n}}+b_{\vec{n}}),
 ~~\tilde{S}^y_{\vec{n}}=i\sqrt{S/2}
 (b^{\dagger}_{\vec{n}}-b_{\vec{n}}),
 ~~\tilde{S}^z_{\vec{n}}=S-b^{\dagger}_{\vec{n}}b_{\vec{n}}$
 with $b^{\dagger}_{\vec{n}}$, $(b_{\vec{n}})$ being the creation
 (destruction) Bose-operator of the spin-excitation on the
 $\vec{n}$-th site and applying the Bogolyubov transformation 
$b_{\vec{n}}=N^{-1/2}\sum\,e^{-i\vec{k}\vec{n}}\left(\cosh(\theta_{\vec{k}
})\,\alpha_{\vec{k}}+
\sinh(\theta_{\vec{k}})\,\alpha^{\dagger}_{-\vec{k}}\right)$ for these
 operators
($\vec{k}$ is the wave vector belonging to the first Brillouin
 zone and $N$ is the number of atoms in the crystal),
the Hamiltonian (\ref{ham}) in the spin-wave approximation 
can be represented
in the form $~H=N
E_{gr}+\sum_{\vec{k}}\omega_{\vec{k}}\,\alpha^{\dagger}_{\vec{k}}\alpha_{\
vec{k}}. $ where \begin{eqnarray} \label{grst}
E_{gr}=E_{cl}-\frac{1}{N}\sum_{\vec{k}}\sinh^2(\theta_{\vec{k}})
\omega_{\vec{k}}\end{eqnarray} is the ground state energy per atom in the
spin-wave approximation. The coefficients of the Bogolyubov transformation
are given by $\tanh^2(\theta_{\vec{k}})=
\left(f_{\vec{k}}-g_{\vec{k}}\right)^2
\left(f_{\vec{k}}+g_{\vec{k}}\right)^{-2}$ where
\begin{eqnarray}\label{FG} f^2_{\vec{k}}=S\,\left(-2J(\vec{q})+
J(\vec{k}-\vec{q})+J(\vec{k}+\vec{q})\right),\nonumber\\
g^2_{\vec{k}}=2S\,\left(J(\vec{k})-J(\vec{q})\right) \end{eqnarray} with
$J({\vec{k}})=\frac{1}{2}\sum_{\vec{a}}J_{\vec{a}}\cos\vec{k}\,\vec{a} $,
and \begin{eqnarray} \label{om} \omega_{\vec{k}}=f_{\vec{k}}\,g_{\vec{k}}
\end{eqnarray} is the magnon energy. $E_{cl}=\,S^2\,J(\vec{q})$ is the 
energy of the spin subsystem per atom in the classical approximation.

The on-site magnetization $M\equiv \langle \tilde{S}^z_{\vec{n}}\rangle=
 S-\langle b^\dagger_{\vec{n}}
 b_{\vec{n}}\rangle$ can be also obtained  by using the Bogolyubov
 transformation. As a result we get
 \begin{eqnarray}
 \label{ons}
 M=S-\frac{1}{N}\sum_{\vec{k}}\sinh^2(\theta_{\vec{k}}).
 \end{eqnarray}

Our goal now is to minimize the  total ground state energy of the system
\begin{eqnarray} \label{en} {\cal F}=\Phi+E_{gr}\equiv E_{gr}+
J\left(\frac{\kappa}{2}(\epsilon^2_1+\epsilon^2_2)- p\epsilon_1\right)
\end{eqnarray} with respect to the variational parameters which are the
components of the deformation tensor  $~~\epsilon_1=\eta(u_{xx}-u_{yy})/2,
~~\epsilon_2=\eta u_{xy}$  and the vector $\vec{q}$. Here the
magneto-elastic constant  $\kappa=4\mu/\left(J\,\eta^2\right)~~$
characterizes  the stiffness of the lattice in terms of the intensity of
the spin-lattice interaction,   $p=(p_x-p_y)/\left(\eta J\right)~~$is the
dimensionless uniaxial pressure. In Eq. (\ref{en}) we omitted the terms of
spin-lattice interaction which correspond to the dilatation (contraction) 
of the lattice ($~(u_{xx}+u_{yy})/2~$) without changing  its symmetry.
These terms don't change the qualitative behavior of the system. The
equations for $\vec{q}$ and $\epsilon_i$ have the form \begin{eqnarray}
\label{meq} \frac{\partial}{\partial \vec{q}} {\cal F}=0,~~~~
\frac{\partial}{\partial \epsilon_i}{\cal F}=0 \end{eqnarray}

Let us consider first the case zero pressure: $p=0$.
In the classical approximation when $S \rightarrow \infty$ the energy of
the system can be represented in the form \begin{eqnarray} \label{clap}
\frac{{\cal F}_{cl}}{J}=S^2 J(\vec{q}) +\frac{\kappa}{2}
(\epsilon^2_1+\epsilon^2_2) \end{eqnarray}

>From Eq.(\ref{clap}) we obtain that the set of
equations (\ref{meq})
has three types of solutions:

i) H-phase: 
$\vec{q}_H=4\pi/3
\left(\cos\phi,\sin\phi\right),~\phi=0,\pm\pi/3.$
The spin structure corresponding to each $\vec{q}_H$ is   a
three-sublattice antiferromaget.Three different  $\vec{q}_H$ represent
three possible antiferromagnetic domains. The lattice structure is an
undistorted triangular lattice $~(\epsilon_1=\epsilon_2=0)~$ with the
group symmetry $D_{6h}$.

ii) L-phase: $\vec{q}_L=2\pi/\sqrt{3}
\left(\sin\phi,\cos\phi\right),~~\phi=0,\pm\pi/3.$
The spin structure corresponding to each $\vec{q}_L$ is a
two-sublattice antiferromagnet. The lattice structure for
$\phi=0$ is determined by the
deformation tensors 
$~\epsilon_1=2 S^2/\kappa,\epsilon_2=0~$
while for two other $\vec{q}_L$  the deformation tensors
can be obtained by  rotations  by $\pm \pi/3$. 
The lattice
is characterized by the group symmetry $D_{2h}$.

iii) S-phase: $\vec{q}_s=\left(q_s,2\pi/\sqrt{3}\right),
\nonumber\\ \epsilon_1=\kappa^{-1}\left(\cos
q_s+\cos(q_s/2)\right),
~\epsilon_2=0$ 
where
$\cos(q_s/2)=\left(\sqrt{9 +32\kappa/S^2}-5 \right)/8$.
Two more vectors $\vec{q}$ and deformation tensors $\epsilon_1$ and
$\epsilon_2$ can be obtained from the above equation by rotations by $\pm
\pi/3$. The spin structure is a spiral antiferromagnet.

We consider the stability of obtained solutions
for the case when
\begin{eqnarray}
\label{subs}
\epsilon_1 \,\geq 0,~~\epsilon_2=0,~~0 \leq q_x \leq \frac{2\pi}{3},
~~q_y=\frac{2\pi}{\sqrt{3}},~ 
\end{eqnarray} 
In the subspace (\ref{subs}) the stability condition
reads
\begin{eqnarray}
\label{stcond}
\frac{\partial^2 {\cal F}}{\partial \epsilon_1^2}\,> 0,~~~
\frac{\partial^2 {\cal F}}{\partial \epsilon_1^2}\,
\frac{\partial^2 {\cal F}}{\partial q_x^2}-
\left(\frac{\partial^2 {\cal F}}{\partial q_x \partial
\epsilon_1}\right)^2\,>\,0
\end{eqnarray} 
One can
obtain from Eqs (\ref{clap}) and (\ref{stcond})  that 
\begin{itemize}
\item   the $L$-phase is stable when $\kappa/S^2 < \frac{9}{4} $
\item   both $H$- and $L$-phases are stable when $\frac{9}{4} < \kappa/S^2
< 5. $ For $\kappa/S^2=4 $ the H- and L-phases have the same energy while
for $\kappa/S^2 < 4 $ the L-phase becomes energetically more favorable.
\item   the $H$-phase is stable when  $\kappa/S^2 > 5 $. \end{itemize} The
spiral S-phase does not correspond to a minimum of the ground state
energy.
  In the subspace  (\ref{subs}) the spiral phase 
for $\frac{9}{4}< \kappa/S^2 < 5 $ corresponds to the saddle point which
separates two stable phases $H$ and $L$.

The wave vectors $\vec{q}_{H}$ and $\vec{q}_{L}$  determine also the
magnetic structure of the H- and L-phases in the spin-wave approximation.
This is not true, however, for the spiral phase where quantum fluctuations
change the value of the spiral wave-vector $q_s$.

\begin{figure}
\centerline{\hbox{
\epsfig{figure=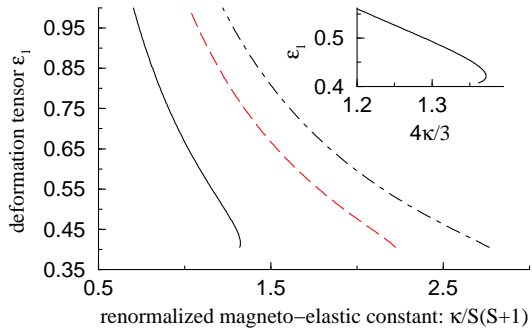,width=75mm,angle=0}}}
\caption{The  deformation tensor 
$\epsilon_1=(u_{xx}-
u_{yy})/2$ \\
for the collinear L-phase (full curve S=1/2,dashed
curve S=1, \\
dotted-dashed curve S=3/2) as a function of the
renormalized \\
magneto-elastic constant $\kappa/S(S+1)$. 
Multistability for \\
the case 
S=1/2 is shown in the inset.}
\label{fig:tr-k-w}
\end{figure}
Evaluating numerically the integrals in the r.h.s. of Eq. 
(\ref{grst}) for $\vec{q}=\vec{q}_H,~\epsilon_1=0$ one can obtain 
that the ground
state energy of the symmetric triangular phase  in the spin-wave
approximation is ${\cal F}_H\approx J\,
\left(-\frac{3}{2}\,S(S+1)+1.1723\,S\right)$. We have also found that the
H-phase is stable for $\kappa >\kappa_H$ where  

\begin{equation}
\label{kappah} 
\frac{\kappa_{H}}{S^2}
= \left\{ \begin{array}{rl} 
                      6.86 \; , & \mbox{for $S=\frac{1}{2}$},   
       \\ 3.85 \; , & \mbox{for $S=1$}, 
       \\ 3.26 \; , & \mbox{for $S=\frac{3}{2}$},
        \\ 2.99 \; , & \mbox{for $S=2$}
                   \end{array} 
              \right. 
\end{equation} 

Considering the low-symmetry L-phase which corresponds to the wave
vector $\vec{q}_L=2\pi(0,1/\sqrt{3})$ we obtain from Eqs
(\ref{FG}), (\ref{om}) that its energy in the
spin-wave  approximation is determined by the expression
\begin{eqnarray}\label{el}      
 \frac{{\cal F}_L}{J}=
-S(S+1)(1+2\epsilon_1)+\frac{\kappa
\epsilon_1^2}{2}+S \,I(\epsilon_1)\nonumber\\
I(\epsilon_1)=(2+\epsilon_1)\frac{4}{\pi^2}\int\limits_{0}^{\frac{\pi}{2}}
\,dy \,  (1- 2\frac{1-\epsilon_1}{2+\epsilon_1}\,\sin^2 y)E(m) 
\end{eqnarray} where  $E(m)$ is the elliptic integral of the second kind
\cite{as} with the parameter  $m=\cos^2 y\,\left(1-
2\frac{1-\epsilon_1}{2+\epsilon_1} \sin^2 y\right)^{-2}$.   One can obtain
from Eq. (\ref{om}) that in the  $\vec{k} \rightarrow 0$ limit the magnon
dispersion in the L-phase has the form $\omega_L(\vec{k})= 2
JS\,(2+\epsilon_1)\sqrt{(5\epsilon_1-2) k_x^2+3 (2+\epsilon_1) k_y^2}$.
This means that the magnons in the low-symmetry L-phase can exist only for
$\epsilon_1 > 0.4$ and only in this interval  we may look for  extrema of
the function ${\cal F}_L$. It is seen from Eqs (\ref{ham})  that the
quantity $J_1=J(1+\epsilon_1/2)$ $\left(J_2=J(1-\epsilon_1)\right)$
corresponds to the antiferromagnetic  exchange between first (second)
neighbors in the spin-wave theory of Heisenberg antiferromagnet on an
anisotropic triangular lattice \cite{trumpo,merino}. Therefore the
interval $1>\epsilon_1 > 0.4$  corresponds to the interval
$~0\,<J_2/J_1\,<\,1/2$ in the spin-wave theory of Heisenberg
antiferromagnet on an anisotropic triangular lattice.

 In Fig. \ref{fig:tr-k-w} we plot the deformation tensor
$\epsilon_1$ which provides an extremum
of ${\cal F}_L$, as a function of the coupling parameter
$\kappa$  for three spin values $S=1/2,1,3/2$.  It is seen 
that the equation $d{\cal F}/d\epsilon_1=0$
 has  solution $~\epsilon_1~$only for $\kappa\,<\,\kappa_{L}$.   
The critical value $\kappa_{L}$ which determines the interval of the
existence of the L-phase depends on the spin $S$. It increases when the
spin $S$ increases, e.g. \begin{equation}\label{kappal}
\frac{\kappa_{L}}{S^2}\,= \left\{ \begin{array}{rl}
                      3.97 \; , & \mbox{for $S=\frac{1}{2}$},  
       \\ 4.47 \; , & \mbox{for $S=1$},
       \\ 4.64 \; , & \mbox{for $S=\frac{3}{2}$},
       \\ 4.73 \; , & \mbox{for $S=2$}
                   \end{array}
              \right.
\end{equation}
but even for $S=4$ it is still less ($\kappa_L/S^2\,\approx 4.75$) than
the critical value for this parameter obtained in the classical
approximation ($\kappa_L/S^2=5$). The reason for this is the strong
contribution to the effective elastic energy of the system from quantum
fluctuations (the second term in the r.h.s. of Eq. (\ref{el})).The quantum
fluctuations change both
 the stiffness of the lattice (the second derivative of the free energy
 (\ref{el}) with respect to the deformation $\epsilon_1$) and the constant
 of the spin-elastic interaction ( the first derivative of the free
 energy). A single-valued monotonic dependence $\epsilon_1(\kappa)$ is
 obtained
for $S \geq 1$. For $S=1/2$ the dependence becomes multi-valued, i.e.
there is an interval of $\kappa$ where two values $\epsilon_1^{(1)}$ and
$\epsilon_1^{(2)}$ ($\epsilon_1^{(1)} < \epsilon_1^{(2)}$ ) of the
deformation tensor correspond to each value of the coupling constant
$\kappa$. Only the solution which corresponds to the deformation
$\epsilon_1^{(2)}$ corresponds to the minimum of the effective elastic
energy (\ref{el}). 

 To find the stability region of the collinear L-phase we numerically
 evaluated integrals in the l.h.s. of the inequality (\ref{stcond}) and
 found that it holds in the interval $\kappa < \kappa_L$. 
  Comparing the stability limits for  H- and L-phases give in Eqs
 (\ref{kappah}) and (\ref{kappal}) we find the following interesting
 results. For $S\,\geq\,1$ we have always $~~\kappa_H \,<\,\kappa_L~~$
 like in the classical approach where the stability regions for H- and
 L-phases overlap. However, for $~S=1/2~$ this is not the case since there
 is an interval $\kappa_L\,<\,\kappa\,<\,\kappa_H$ where neither the L-
 nor the H-phase exist.

 The nature of the phase in the interval $(\kappa_L,\kappa_H)$ or in other
 words for $\epsilon_1 < 0.4$, cannot be clarified in the framework of the
 spin-wave approach. The reason why this approach fails is the following.
 As it was mentioned above the vectors $\vec{q}_s$ obtained in the
 classical approach are not solutions of the extrema conditions
 (\ref{meq}) in the spin-wave approximation. On the other hand, as it is
 seen from Eqs (\ref{om}) and
(\ref{FG}) the  magnon frequency $\omega_{\vec{k}}$ is real   in the
interval
  $\epsilon_1 < 0.4$ only  for $\vec{q}$ obtained in the classical
  approach. 
For other $\vec{q}$ values    $~\omega_{\vec{k}}^2~$ is not positive
definite. Thus to check the stability of  the   corresponding phase is
stable  one should calculate the magnon dispersion    beyond the spin-wave
 approach.
\begin{figure}
\centerline{\hbox{
\epsfig{figure=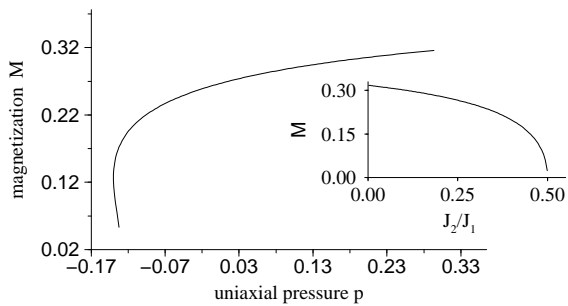,width=75mm,angle=0}}}
\caption{The effect of the uniaxial pressure
on the on-site\\
 magnetization in the L-phase 
($S=1/2,\kappa=1.315$). The \\
inset represents the 
magnetization as a function of the \\
ratio 
 $J_2/J_1=2\,(1-\epsilon_1)/(2+\epsilon_1)$ of the effective 
 exchange \\
 integrals between the second and
first neighbors in the \\
presence of the uniaxial pressure.}
\label{fig:tr-m-j2j1}
\end{figure}
   Let us consider how the L-phase evolves in the presence of an uniaxial
   pressure $p$. From Eqs (\ref{en}) and (\ref{meq}) we get
   \begin{eqnarray} \label{eqp}
   \vec{q}=\frac{2\pi}{\sqrt{3}}\,\left(0,1\right),~~~\epsilon_2=0,\nonumber\\
   p=\kappa\epsilon_1-2S(S+1)+S\,\frac{d I(\epsilon_1)}{d
\epsilon_1}
\end{eqnarray}
 In Fig. \ref{fig:tr-m-j2j1} we show results for the on-site magnetization
  in the L-phase for $S=\frac{1}{2}$ as a
function   of the uniaxial pressure $p$. Fig. 2 and in particular
 its inset shows that our results are qualitatively different from those
 of a recent spatially anisotropic triangular lattice study
 \cite{trumpo,merino}. In contrast to the $(J_1,J_2)$ -model
 \cite{trumpo,merino} where the ratio $J_2/J_1$ was a free parameter in
 our model the corresponding parameter in a compressible triangular
 lattice $J_2/J_1=2\,(1-\epsilon_1)/(2+\epsilon_1)$ is determined
 self-consistently from the extrema conditions (\ref{meq}). It is seen
 that in the interval of the existence of the L-phase we have $J_2/J_1 <
 1/2~~$ $\,(\epsilon_1 > 0.4~$) and the on-site magnetization is finite.
 It is interesting that there exists an interval where two values of the
 magnetization $M$ correspond to each value of the  pressure $p$. One can
 show, however, that only the state with a larger value of  $M$ is stable.

  We conclude that  the Heisenberg antiferromagnet on an compressible
  triangular lattice differs qualitatively from the Heisenberg
  antiferromagnet on an anisotropic triangular lattice. Self-consistent
  treatment of the elastic degrees of freedom shows that the spiral
  magnetic phase is unstable in the classical approximation. This 
  approach shows also that in the spin-wave approximation the interaction
  between quantum fluctuations and elastic degrees of freedom stabilizes
  the collinear L-phase.

 {\bf Acknowledgements}

Yu. Gaididei is grateful for the hospitality of the University of Bayreuth

 where this work was performed. Partial support from the DRL grant Nr.:
 UKR-002-99 is also acknowledged.

\end{multicols}
\end{document}